\documentclass[aps, prd, 10pt, twocolumn, nofootinbib, superscriptaddress, elide, sort&compress, preprintnumbers]{revtex4-1}
\usepackage[colorlinks=true, citecolor=blue, urlcolor=blue, linkcolor=blue, breaklinks=true, pdfpagelabels=false]{hyperref}
\usepackage[utf8]{inputenc}
\usepackage{lmodern,amsmath,slashed,color,graphicx,feynmp}

\allowdisplaybreaks
\makeatletter\g@addto@macro\bfseries{\boldmath}\makeatother

\def\equationautorefname~#1\null{Eq.\,(#1)\null}
\def\pageautorefname\nobreakspace{p.}

\makeatletter\renewcommand{\p@subsection}{\thesection.}\makeatother%

\DeclareMathOperator{\dif}{d}
\renewcommand{\arraystretch}{1.5}

\newcommand{\beq}{\begin{equation}}
\newcommand{\eeq}{\end{equation}}

\begin{document}

\DeclareGraphicsRule{*}{mps}{*}{}
\setlength{\unitlength}{1mm}
\fmfcmd{%
prologues:=3;
thin := .5pt;
arrow_ang := 12;
arrow_len := 3.5thick;
}

\newcommand{\desy}{Deutsches Elektronen Synchrotron (DESY),
Notkestraße 85, D–22607 Hamburg, Germany}
\newcommand{\cornell}{Laboratory for Elementary Particle Physics, Cornell
University, Ithaca, NY 14853, USA}
\newcommand{\louvain}{Centre for Cosmology, Particle Physics and Phenomenology,\\
Université catholique de Louvain, B-1348 Louvain-la-Neuve, Belgium}
\newcommand{\mainz}{PRISMA Cluster of Excellence \& Mainz Institute for
Theoretical Physics,\\Johannes Gutenberg University, 55099 Mainz, Germany}
\newcommand{\CU}{Department of Physics, University of Calcutta, 92 A. P. C.
Road, Kolkata 700 009, India}

\author{Gauthier Durieux}	\affiliation{\desy}\affiliation{\cornell}\affiliation{\louvain}
\author{Yuval Grossman}		\affiliation{\cornell}
\author{Matthias K\"onig}	\affiliation{\mainz}
\author{Eric Kuflik}		\affiliation{\cornell}
\author{Shamayita Ray}		\affiliation{\cornell}\affiliation{\CU}

\preprint{DESY 15-235}
\preprint{MITP/15-111}

\title{Rare \texorpdfstring{$Z$}{Z} decays and neutrino flavor universality}

\begin{abstract}
We study rare four-body decays of the $Z$-boson involving at least one neutrino and one charged lepton. Large destructive interferences make these decays  very
sensitive to the $Z$ couplings to neutrinos.  As the identified charged leptons can determine the neutrino flavors, these decays probe the universality of the $Z$ couplings to neutrinos.  The rare four-body processes could be accurately measured at future lepton colliders, leading to percent level precision.
\end{abstract}

\maketitle

\section{Introduction}
In the 1990s, the electron-positron colliders working at the $Z$ pole, LEP and
SLC, provided critical tests of the Standard Model (SM). The precision
measurements of many electroweak observables still set strong bounds on physics
beyond the SM. Future circular lepton colliders, like the Circular Electron
Positron Collider (CEPC) or Future Circular Collider (FCC-ee, formerly TLEP),
with anticipated yields of about $10^{12}$ $Z$-bosons~\cite{Zimmermann:2014qxa},
could test the SM further by measuring rarer $Z$ decays.

In this work, we focus on the four-body $Z\to~jj\,l\nu_l$, $Z\to l\,l\nu\nu$,
and $Z\to l\,l^{\prime}\nu_l\nu_{l^{\prime}}$ decays, where $j$ stands for a
jet. In the SM, these rather clean decays have branching fractions of order of
$10^{-8}$. Future colliders could bring the precision on their measurements down
to the one-percent level. In the first and third channels listed above, the
neutrino flavor matches the charged lepton flavor, allowing for identification of
the neutrino flavor. Separate sensitivity to the couplings of the $Z$-boson to
each flavor of neutrino is therefore obtained.

Instead of relying on a specific new-physics scenario, we adopt a simplified
approach and study the dependence of the decay rates on the relevant couplings.
This allows for the identification of the important interference effects which lead to
sensitivities comparable to, or better than, the ones obtained from other
processes. Existing constraints on the flavor universality of the neutrino
neutral currents derive from various other sources and are discussed in detail
below. We stress that each observable depends on a different combination of
couplings, and that the complementarity of several observables needs to be
exploited for constraining the couplings individually. 
The $Z$ decays we study constitute a fairly direct probe as their dependence on the couplings
is often rather simple. For instance, each $Z\to~jj\,l\nu_l$ decay rate probes
directly one single coupling of the $Z$ to neutrinos, while two of them
enter the $Z\to l\,l^{\prime}\nu_l\nu_{l^{\prime}}$ rate.

While our study is done in a model-independent way, it is still relevant to ask
how deviations from the SM could be generated. Since the universality of the
couplings of the $Z$ to neutrinos is a consequence of gauge symmetry, it rests
on rather robust theoretical grounds. To some extent, it is constrained
experimentally by the observed universality of the $Z$ couplings to charged
leptons. New-physics scenarios featuring mixings of the $Z$ or neutrinos to new
states could alter the $Z$ couplings to neutrinos. {The
exploration of such models} would require dedicated studies that are beyond the
scope of the current paper. Thus, the $Z$ decays considered here can be
thought of in two ways. First, taking into account our theoretical
model-building philosophy, these modes could be treated as ``SM candles'' to be
compared with other probes, like those coming from neutrino experiments. On the
other hand, we can look at these decays as a probe of unknown physics. We stress that these
decays should be measured experimentally, regardless of theoretical prejudices. 

Experimental searches for $Z$ decays to similar final states have been carried
out at LEP~\cite{Adriani:1992pq, Abreu:1996pa}. They focused on specific
kinematical regions (with a displaced secondary vertex, or a boosted subsystem)
that are populated in the presence of massive sterile neutrinos. A study of the
probing power of the FCC-ee on such scenarios has been presented in
Refs.~\cite{Blondel:2014bra,Antusch:2015mia}.

\section{Existing bounds} \label{sec:existingbounds}

In this section, we set up the framework for studying neutrino-$Z$ couplings and
review the existing constraints.  

\subsection{Notations}

For the purpose of showcasing the potential of the rare $Z\to jj\,l\nu_l$, $Z\to
l\,l\nu\nu$, and $Z\to l\,l^\prime\nu_l\nu_{l^\prime}$ decays to measure the
$Z$-boson couplings to neutrinos, and to compare to existing experimental
constraints, we consider minimal modifications of the SM interactions of the
$Z$-boson to neutrinos, rescaling them by a real number,
\begin{equation}
{\cal L}_{Z\nu\nu} = -\sum_{l=e,\mu,\tau}
	C_{\nu_l}				\;
		\frac{g}{2 \cos{\theta_W}}	\;
		\bar\nu_l \gamma_\rho P_L \nu_l	\; Z^\rho
	\; .
\label{nsi_C}
\end{equation}
The SM is recovered when $C_{\nu_l}=1$. Note that we set the lepton flavor
violating off-diagonal couplings to zero, as is the case in the SM. These
couplings are highly constrained, for instance, by neutrino oscillations in
matter, as will be discussed in \autoref{sec:osc}. The Hermiticity of the
Lagrangian allows phases to be present only in the flavor off-diagonal
couplings.
Only the known left-handed neutrinos are considered. Therefore, the Lagrangian
introduces three new parameters: $C_{\nu_e}$, $C_{\nu_\mu}$, and $C_{\nu_\tau}$.

We next connect our notations to those used in the context of neutrino
experiments. Earlier studies of the nonstandard interactions (NSIs) of
neutrinos in neutrino scattering and neutrino oscillation experiments typically
used an effective Lagrangian relevant at energies much below the $Z$ mass:
\begin{eqnarray}
	{\cal L}^{\rm eff} = 
	{\cal L}_{\rm SM}^{\rm eff}
		-\epsilon_{\alpha \beta}^{f M} 2\sqrt{2} G_F		
				\left( \bar\nu_\alpha \gamma_\rho P_L \nu_\beta\right)
		\left( \bar f \gamma^\rho P_M f\right),
\label{nsi_epsa}
\end{eqnarray}
where $\alpha,\beta = e, \mu, \tau$, $f = e, u, d$ and $M = L,R$. 
Integrating out the $Z$-boson in $\mathcal{L}_\text{SM} +
\mathcal{L}_{Z\nu\nu}$, the correspondence between the parameters of
\autoref{nsi_C} and \autoref{nsi_epsa} is
\begin{equation}
	\epsilon_{\alpha \alpha}^{f M} = g^{f}_M (C_{\nu_\alpha}-1),
\end{equation}
where
\begin{equation}\def\arraystretch{2.0}
	\begin{array}{l@{\qquad}l}
	g^{\ell}_L = -\frac{1}{2}+ \sin^2\theta_w,		&
	g^{\ell}_R = \sin^2\theta_w,				\\
	g^{u}_L =  \frac{1}{2}- \frac{2}{3}\sin^2\theta_w,	&
	g^{u}_R = -\frac{2}{3} \sin^2\theta_w,			\\
	g^{d}_L = -\frac{1}{2}+ \frac{1}{3}\sin^2\theta_w,	&
	g^{d}_R =  \frac{1}{3}\sin^2\theta_w.
	\end{array}
	\label{epsa-g-C}
\end{equation}

\subsection{\texorpdfstring{$Z$}{Z} pole data}

We now review some of the more stringent bounds on the $C_{\nu_l}$ couplings.
First, the measurement of the total and visible width of the $Z$ constrains the number of
neutrinos~\cite{ALEPH:2005ab}:
\beq
	N_\nu = 2.984 \pm 0.008.
\eeq
In the parametrization of \autoref{nsi_C},
\beq
	N_\nu =	C_{\nu_e   }^2+
		C_{\nu_\mu }^2+
		C_{\nu_\tau}^2,
\eeq
and the measurement translates into a tight bound on the sum of the squared
$C_{\nu_l}$'s. Note that the invisible width of the $Z$ does not probe the
coupling of each neutrino species separately, and is insensitive of the
sign of the couplings.

\begin{figure}
	\includegraphics[width=\columnwidth]{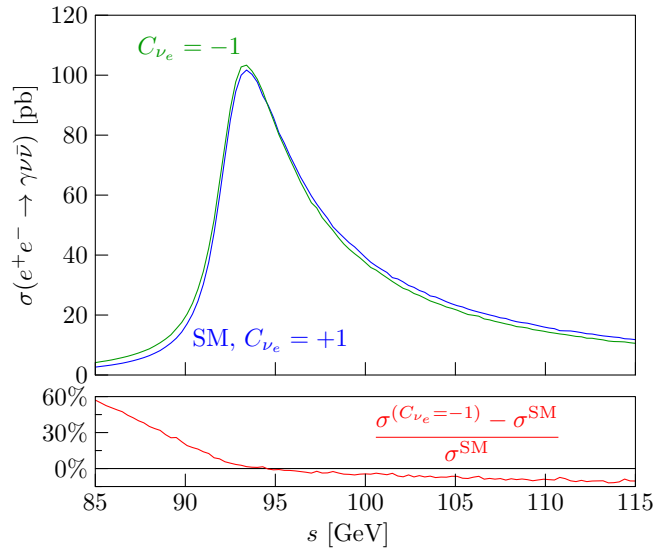}
	\caption{Cross-section of the $e^+e^-\to \gamma\nu\bar\nu$ process, as
	         a function of the center-of-mass energy, for $C_{\nu_e}=+1$ and
	         $C_{\nu_e}=-1$. Relative difference between the two
	         distributions obtained with
	         \textsc{MadGraph5}~\cite{Alwall:2014hca} at leading order for
	         $E_\gamma > 1$~GeV and $45^\circ < \theta_\gamma < 135^\circ $. }
	\label{fig:ee2avv}
\end{figure}

The $e^+e^-\to \gamma\nu\bar\nu$ cross-section was also used as a neutrino
counting observable, giving $N_\nu = 2.92 \pm 0.05$~\cite{Agashe:2014kda}. Off
the cross-section peak there is sizable interference between the $s$-channel $Z$
and $t$-channel $W$ exchange contributions. It produces a linear dependence of
the cross-section on $C_{\nu_e}$. We estimate the relative difference in
cross-sections for $C_{\nu_e}=+1$ and $C_{\nu_e}=-1$ to be of the order of
$10\%$ near the $Z$ pole (see \autoref{fig:ee2avv}) and conclude that the
possibility of a negative $C_{\nu_e}=-1$ is untenable.  

Changes in the way the $Z$ couples to neutrinos can also affect the $Z\to l^+
l^-$ partial widths at the one-loop level. Without a gauge invariant framework,
one cannot derive reliable bounds on the $C_{\nu}$ couplings. Using naive
dimensional analysis, the correction to the $Z\to l^+ l^-$ is expected to be
around ${C_{\nu_l}}/{16 \pi^2}$. Given the current experimental precision on the
$Z\to l^+ l^-$ rates~\cite{Agashe:2014kda}, the naive expectation is that a
bound of $|C_{\nu_l}-1| \sim 0.1$ can be obtained within specific models. This
is in the same range as other existing bounds discussed in this section.
In a specific model, however, and with the future experimental precision, these
effects may become important.

\subsection{Neutrino scattering data}
\label{bound-scattering}

Neutrino scatting experiments provide strong constraints on neutrino
interactions with electrons and quarks of the first generation. They utilize
neutrino beams of known flux and flavor. 
Combined bounds can be found in
Refs.~\cite{Barger:1991ae,Berezhiani:2001rs,Davidson:2003ha,Escrihuela:2011cf}, including the
measurements of the following processes:
(i) Electron-neutrino scattering off electrons, $\sigma_{\nu_e e \to \nu e}$, 
by LSND~\cite{Auerbach:2001wg}, 
(ii) the ratio of neutral-current to
charged-current scatterings of electron-neutrino off nucleons,
$\sigma_{\nu_e N \to \nu X}/\sigma_{\nu_e q \to e X}$, by
CHARM~\cite{Dorenbosch:1986tb}, 
(iii) muon-neutrino scattering off
elections, $\sigma_{\nu_\mu e \to \nu e}$, by CHARM
II~\cite{Vilain:1994qy}, and (iv) the ratios of neutral-current to
charged-current scatterings of muon-neutrino off nucleons,
$\sigma_{\nu_\mu N \to \nu X}/\sigma_{\nu_\mu q \to \mu X}$, by
NuTeV~\cite{Zeller:2001hh}.
Translating the bounds from $\epsilon_{\alpha \alpha}^{f M}$ to $C_{\nu_l}$ gives
\begin{equation}
	0.72 < C_{\nu_e}   < 1.32\, , \qquad
	0.99 < |C_{\nu_\mu}| < 1.01\,,
\label{C1}
\end{equation}
with no constraint on $C_{\nu_\tau}$. The interference
between charged and neutral currents provides a handle on the sign of
$C_{\nu_e}$ and excludes, here again, a negative value.

Global fits of the electron-neutrino NSI parameters to scattering rate measurements
were also performed in Refs.~\cite{Barranco:2005ps, Barranco:2007ej,Forero:2011zz}. Their
inclusion of additional data from MUMU~\cite{Daraktchieva:2003dr},
Rovno~\cite{Derbin:1986jh}, LAMPF~\cite{Allen:1992qe},
Krasnoyarsk~\cite{Vidyakin:1992nf} and Texono~\cite{Deniz:2009mu} resulted in
tighter bound on $C_{\nu_e}$:
\begin{eqnarray}
	0.94 < C_{\nu_e} < 1.07.
	\label{C2}
\end{eqnarray}
Note that the constraints on NSI parameters in the above analyses derived from
multidimensional fits for the $\epsilon_{\alpha \alpha}^{f P}$ degrees of
freedom. Each $C_{\nu_l}$ represents only one degree of  freedom in the
$\chi^2$-distribution. We did not correct for this in \autoref{C1} and
\autoref{C2} and consider these bounds as estimates to be compared with the
prospects at future colliders.

\subsection{Neutrino oscillations \label{sec:osc}}

As discussed in Refs.~\cite{Davidson:2003ha, Bolanos:2008km, Biggio:2009nt,Escrihuela:2009up, Miranda:2015dra}, 
both flavor-diagonal and
flavor-changing NSI parameters are constrained by neutrino oscillation
experiments. They can affect neutrino production, detection, and propagation through
matter. 

Atmospheric neutrinos are very sensitive to matter NSIs as they travel a long
distance through the Earth. Since the Earth is made up of approximately
equal number of protons, neutrons and electrons, the atmospheric neutrino oscillation 
experiments bound the quantity
\begin{equation}
\epsilon^\oplus_{\alpha \beta} =
	\sum_M \left(
	  3\epsilon^{u M}_{\alpha \beta}
	+ 3\epsilon^{d M}_{\alpha \beta}
	+ \epsilon^{e M}_{\alpha \beta}
	\right) \; .
\label{epsa-atm}
\end{equation}
Experimental studies of the matter NSIs with the Super-Kamiokande atmospheric
neutrino data~\cite{Ohlsson:2012kf, Mitsuka:2011ty} result in
\begin{equation}
\left| \epsilon_{\tau\tau}^\oplus - \epsilon_{\mu\mu}^\oplus \right|  < 0.147 \; ,
\end{equation}
where the analysis in Ref.~\cite{Mitsuka:2011ty} assumed that
neutrinos interact only with the $d$-quarks inside the
Earth to fix the normalization~\cite{GonzalezGarcia:1998hj}. Considering the complete Earth contribution,
as given in \autoref{epsa-atm}, and using \autoref{epsa-g-C}, the bound on
$C_{\nu_l}$ becomes
\begin{eqnarray}
\left|  C_{\nu_\tau}  - C_{\nu_\mu }\right| < 0.294 \; ,
\end{eqnarray}
which is much weaker than the constraints on $C_{\nu_l}$ discussed in
\autoref{bound-scattering}.

In our study we do not consider flavor-changing NSI parameters, but briefly
discuss the bounds from neutrino oscillations. Super-Kamiokande atmospheric
neutrino data yields
\begin{equation}
\epsilon_{\mu\tau}^\oplus < 0.033\; ,
\end{equation} 
which was obtained with a normalization that assumes neutrinos
interact only with $d$-quarks~\cite{Ohlsson:2012kf, Mitsuka:2011ty} inside the Earth.
While earlier studies~\cite{Kopp:2007ne,Ohlsson:2008gx,Leitner:2011aa} 
demanded that reactor experiments bound the NSI
parameters $\epsilon_{e\mu}$ and $\epsilon_{e\tau}$, recent 
analysis~\cite{Girardi:2014gna,Agarwalla:2014bsa} showed that Daya Bay 
experiment cannot constrain the flavor-changing NSI parameters because 
 of their strong correlation with the reactor angle $\theta_{13}$.
It can, however, bound $|\epsilon_{ee}| < {\cal O}(10^{-2})$, 
when the parameter is considered to be real, and a normalization error in the 
neutrino flux is taken into account. No bound can be put if an arbitrary phase is allowed.
The accelerator neutrino experiments like K2K~\cite{Friedland:2005vy},
MINOS~\cite{Kitazawa:2006iq,Kopp:2010qt, Mann:2010jz, Friedland:2006pi,
Blennow:2007pu, Isvan:2011fa}, T2K~\cite{Coelho:2012bp}, OPERA~\cite{Ota:2002na,
EstebanPretel:2008qi, Blennow:2008ym} and NO$\nu$A~\cite{Friedland:2012tq} also
bound the flavor-changing NSI parameter $\epsilon_{e\tau}$. 
 Future neutrino-factory
experiments could test the off-diagonal NSI parameters down to the $10^{-3}$
level, whereas diagonal NSI parameter combinations such as $(\epsilon_{ee} -
\epsilon_{\tau\tau})$ and $(\epsilon_{\mu\mu} - \epsilon_{\tau\tau})$ could only
be tested down to $10^{-1}$ and $10^{-2}$, respectively~\cite{Coloma:2011rq}.

\section{Interference patterns}\label{sec:threeBody}
Before turning to the four-body decays of interest and their sensitivity to the
$C_{\nu_l}$ couplings, we detail the interference pattern of the simpler
three-body $Z\to Wl\nu_l$ decays.

They receive three contributions, respectively proportional to the $Z$ couplings
to neutrinos, to charged leptons, and to the $W$-boson. The corresponding
diagrams are shown in \autoref{fig:diagramThreeBody}. From the full analytical
expressions given in the \hyperref[app:calcAmpl]{Appendix}, we obtain the
tree-level SM branching fraction:
\begin{equation}
	\Gamma^\text{SM}(Z\to W l\, \nu_l) \simeq 1.99\times 10^{-8}~\text{GeV}
\end{equation}
for each lepton flavor. Given the total $Z$-boson width of $2.50$~GeV, this
corresponds to a branching ratio of the order of $10^{-8}$. The rates for
distinct lepton flavors differ by ratios of the charged lepton masses to the
$Z$-boson mass, which we have neglected in this section.

To examine the pattern of interferences, we momentarily introduce rescaling
parameters for the $Z$ couplings to the charged leptons and $W$-boson, $C_{l}$
and $C_{W}$, respectively. We assume that these parameters are real, as is the case for 
the $C_{\nu_l}$'s. The total decay rate is then:
\begin{multline}
	\frac{\Gamma(Z\to W l \nu_l)}{10^{-8}\text{ GeV}}
	\simeq \\
	\left(\begin{array}{@{}c@{}}
			C_{\nu_l}\\
			C_l\\
			C_W
			\end{array}
		\right)^{\hspace{-.7ex}T} 
		\left(
			\begin{array}{@{}ccc@{}}
			1.36 & -0.24 & -1.59 \\
			-0.24 & 0.39 & -0.86 \\
			-1.59 & -0.86& 5.63 \\
			\end{array}
		\right) 
		\left(
			\begin{array}{@{}c@{}}
			C_{\nu_l}\\
			C_l\\
			C_W
			\end{array}
		\right)\,.
\label{eq:decRate3}
\end{multline}
Since we consider the $W$ as a final-state particle, we have neglected its width
here.
 
The coupling strengths of the $Z$ to neutrinos, charged leptons, and $W$ mainly
determine the magnitude of each contribution taken in isolation. Remarkably, all
their interferences are destructive and correct the decay rate by a factor of
$3.7$. Interferences linear in the coupling of the $Z$ to neutrinos alone are
responsible for a factor of $2.0$. Flipping the sign of $C_{\nu_l}$ would
increase the partial decay rate by a factor of $4.7$. The rate of the $Z\to W
l\nu_l$ decay has a high sensitivity to modifications of its magnitude and sign,
for each flavor $l$, individually. Focusing on the neutrino couplings, fixing
$C_l=C_W=1$, the three-body width is:
\beq
	\frac{\Gamma(Z\to W l\, \nu_l)}{10^{-8} \rm~ GeV} = 
	  4.3
	- 3.7\, C_{\nu_l}
	+ 1.4\, C_{\nu_l}^2.
\eeq

\begin{figure}[t!]
\centering
\includegraphics[width=.45\textwidth]{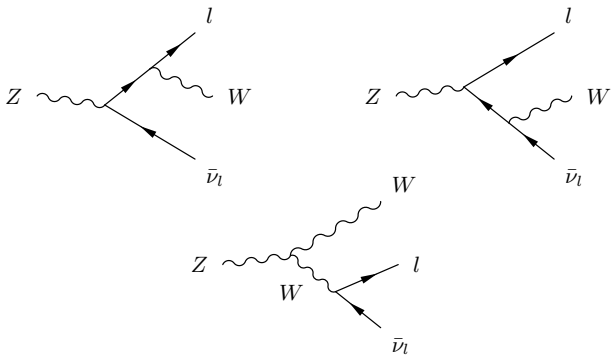}
\caption{Tree-level diagrams contributing to $Z\to Wl\nu_l$ decays.}
\label{fig:diagramThreeBody}
\end{figure}%

\section{Four-body decays}\label{sec:fourBody}

\begin{figure}[t!]
\centering
\includegraphics[width=.4\textwidth]{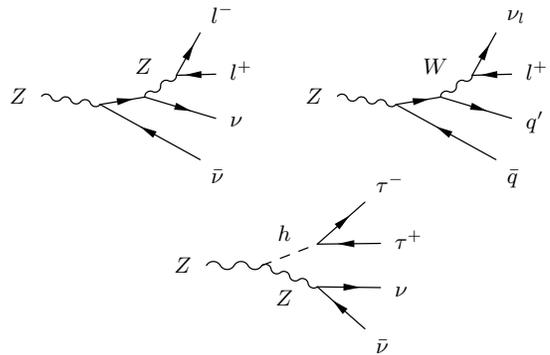}
\caption{Some contributions to the four-body $Z\to jjl\nu_l$ and $Z\to
         l\,l\nu\bar\nu$ decays that do not proceed through a $Wl\nu_l$
         intermediate state. }
\label{fig:diag4}
\end{figure}

In practice, the observable processes are four-body decays. They receive
contributions from diagrams featuring an intermediate $W\,l\,\nu_l$ state as well as
new contributions that do not derive from the three-body process discussed in
the previous section (see \autoref{fig:diag4}).  Higher-order
corrections could be important given the future experimental accuracy. In this
first qualitative study, we only show leading-order results.

We focus on three different channels. In the semileptonic $Z\to jjl\,\nu_l$ decay,
the flavor of the neutrino is fixed by that of the lepton and the coupling of
each neutrino to the $Z$ can be probed separately.
Among the four-body decays that are sensitive to the $Z$ neutrino couplings, it
also has the highest rate. The fully leptonic $Z\to l\,l'\nu_l\nu_{l'}$ decay
involves two leptons of distinct flavors. In that case, interferences between
diagrams where the $Z$ couples to neutrinos of different flavors render the
analysis more involved. On the other hand, in $Z\to l\,l\nu\nu$, with two leptons of the same flavor, 
all species of neutrinos can be produced irrespectively of the flavor of the
charged leptons (see, e.g., the first diagram of \autoref{fig:diag4}). The
presence of two couplings of the $Z$ to neutrinos in the corresponding diagrams
also introduces cubic and quartic $C_{\nu_l}$ dependences in the decay rate.
However, there are no interferences proportional to two different
$C_{\nu_l}$'s in this third channel. 

\begin{figure*}[ht!]
\includegraphics[width=.33\textwidth]{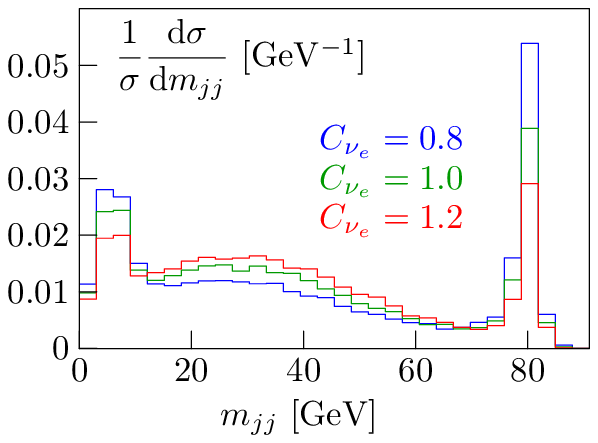}%
\includegraphics[width=.33\textwidth]{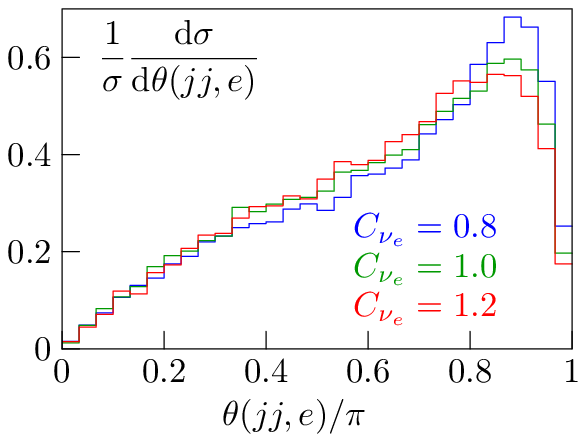}%
\includegraphics[width=.33\textwidth]{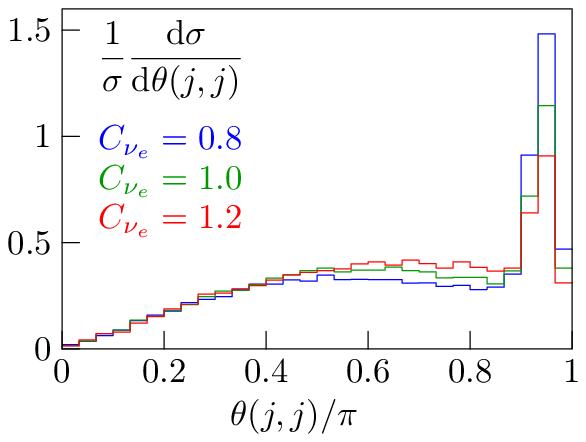}%
\caption{The effect of positive $C_{\nu_l}$ variations on some differential
distributions in $Z\to jjl\nu_l$. The dijet invariant mass, angle between the
dijet system and charged lepton, and angle between the two jets are displayed.}
\label{fig:had_diff}
\end{figure*}

Using \textsc{MadGraph5}~\cite{Alwall:2014hca}, we extract the dependence of
each decay rate on the $C_{\nu_l}$ coefficients. These are given in
Eqs.~(\ref{llnunu}--\ref{emununu}) of the \hyperref[app:fourbody]{Appendix}. The
$\tau$ mass has been kept nonvanishing and marginally affects some of the
numerical factors. Because of additional contributions to the four-body decays
and of the small phase space available in the three-body decay (which requires an
on-shell $W$), the rate of the four-body processes are much higher than what
would have been obtained by using a narrow-width approximation on the three-body
decays. Fixing $C_{\nu_l}=1$, the SM decay widths are
\begin{eqnarray*}
\frac{\Gamma^\text{SM}(Z\to l\,l\nu\bar\nu)}{10^{-8}\text{ GeV}}
	&\simeq& \left\{
		\begin{array}{l@{\;\text{ for }}l}
		2.4 & l=e,\mu\\
		2.3 & l=\tau
		\end{array}\right.
\\[1mm]
\frac{\Gamma^\text{SM}(Z\to l\,\nu_l j j)}{10^{-8}\text{ GeV}}
	&\simeq& \left\{
		\begin{array}{l@{\;\text{ for }}l}
		6.5 & l=e,\mu\\
		6.3 & l=\tau
		\end{array}\right.
\\[1mm]
\frac{\Gamma^\text{SM}(Z\to l\:l'\nu_{l} \nu_{l'})}{10^{-8}\text{ GeV}}
	&\simeq& \left\{
		\begin{array}{l@{\;\text{ for }}ll}
		1.5 & l=e,     & l'=\mu\\
		1.4 & l=e,\mu, & l'=\tau
		\end{array}\right.
\end{eqnarray*}
for each lepton charge assignment.

\begin{figure}[t!]
\includegraphics[width=\columnwidth]{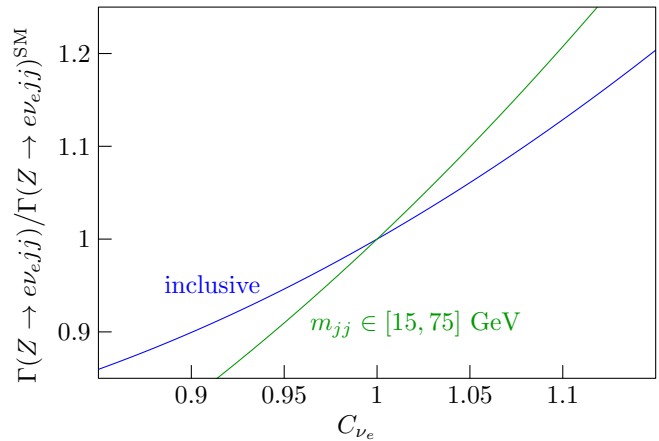}
\caption{Improvement obtained in the sensitivity to positive $C_{\nu_e}$ when a
         $m_{jj}\in [15,75]$~GeV cut is imposed on the dijet invariant mass of
         the $Z\to jje\nu_e$ decay. Similar results are expected for $Z \to \mu
         \nu_\mu jj$ and $Z \to \tau \nu_\tau jj$.}
\label{fig:had_cut}
\end{figure}

The $Z\to jj\,l\nu_l$ process has the highest rate and the simplest dependence
in the $C_{\nu_l}$ couplings. When $C_{\nu_l}=-1$, the destructive interferences
discussed in \autoref{sec:threeBody} cause a dramatic increase of the width by
a factor of $4.1$ (to $27\times 10^{-8}$~GeV for $l=e,\mu$, and $26\times
10^{-8}$~GeV for $l=\tau$). The changes induced by positive $C_{\nu_l}$'s in
differential distributions of $Z\to jj\,l\,\nu_l$ are moderate (see
\autoref{fig:had_diff}). An increased sensitivity could be obtained by selecting
dijet invariant masses in the $[15,75]$~GeV interval (see
\autoref{fig:had_cut}). Regions of the phase space with enhanced sensitivities
could also be studied and exploited in the channels involving two final-state
charged leptons.

\begin{figure}[h!]
\centering
\includegraphics[width=.9\columnwidth]{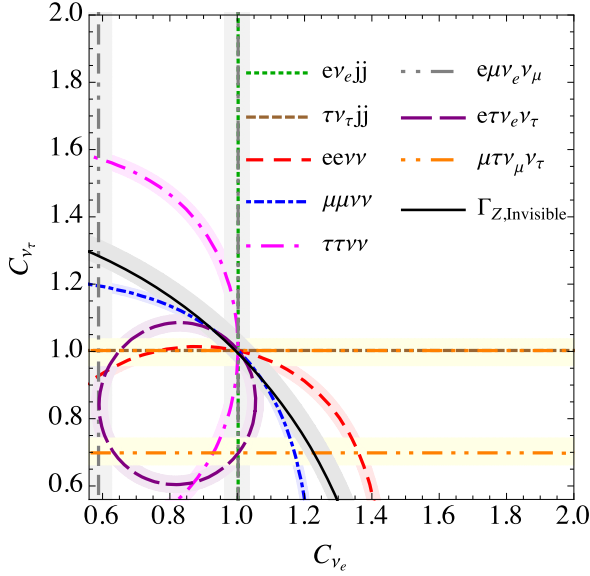}
\caption{Bounds on $C_{\nu_e}$ and $C_{\nu_\tau}$ from all the relevant decay
         channels as listed in Eqs.~(\ref{llnunu}--\ref{emununu}), 
         with $C_{\nu_\mu} = C_{\nu_l}^{\rm SM} = 1$, assuming 2\% uncertainty 
         in the rate measurement.} 
         \label{fig:contours}
\end{figure}

In the $C_{\nu_e}-C_{\nu_\tau}$ plane of \autoref{fig:contours}, we display the
lines along which several relevant partial decay widths take their SM values.
The muon-neutrino coupling to the $Z$ has been fixed to its standard-model value
$C_{\nu_\mu}^{\rm SM}=1$. The bands around each line assume $2\%$ uncertainties on
the rate measurements, a precision that should be achieved at future colliders.
The current bound on $\sum_l C_{\nu_l}^2$ from the total $Z$-boson width is also
displayed. It should be noted that $Z
\to \mu \mu \nu \nu$ has a strong dependence on $C_{\nu_\mu}$ and the band is
expected to get broadened when allowing $C_{\nu_\mu}$ to vary within its allowed
range. A combination of several channels would  bring the constrains on
the $C_{\nu_e}$ and $C_{\nu_\tau}$ couplings down to the percent level at which
the magnitude of $C_{\nu_\mu}$ is currently known.  A
negative value for the latter could be unambiguously excluded. Only a
tiny volume of the full $C_{\nu_e}-C_{\nu_\tau}-C_{\nu_\mu}$ parameter space
would remain allowed if no deviation is observed.

\section{Conclusions}

We have demonstrated the high and differentiated sensitivities of certain
four-body decays of the $Z$ to its couplings to each flavor of neutrino. They
are sourced by large destructive interferences. While, in our study, we
concentrated on the coupling of the $Z$ to neutrinos, deviations from the SM
could also occur in several other couplings that enter the decays discussed. So,
we emphasize that nonstandard interactions should be probed in several
independent ways. We expect that future circular colliders running at the $Z$
peak will measure the suggested decays at the one-percent level. Future neutrino
scattering and oscillation experiments will also further probe the low-energy
limit of interactions that depend on the same couplings of the $Z$ to neutrinos.
It is the combination of these experiments that will give the strongest probing
power and ensure the robustness of the obtained limits.

There are several possible outcomes to such a program. These experiments may
agree with SM predictions, and set stronger
bounds on the deviations of the couplings from their SM values. Alternatively,
some deviations might be established. In that case, a combination of experiments
should be used to identify unambiguously their origin. We can imagine a situation
in which neutrino oscillation observations deviate from the SM expectations
while $Z$ decays rates agree with them. That could be an indication of a new
heavy mediator of neutrino interactions. Another possible scenario could be that
of a deviation only found in the four-body decay of the $Z$ involving two
charged leptons of identical flavor, but neither in that featuring jets, nor in that
involving charged leptons of different flavors. Such an outcome could be
explained by a new source of a triple-$Z$ vertex.

\subsection*{Acknowledgments}
We thank Jakub Scholtz for useful discussions. EK thanks Josh Ruderman for initial discussion on this project. 
The work of GD was supported in part by the FNRS, Belgium and by the Belgian American Education 
Foundation, USA. The work of YG and EK is supported in part
by the U.S.\ National Science Foundation through grant PHY-1316222 and by a
grant from the BSF. MK is supported by the DFG Graduate School Symmetry Breaking
in Fundamental Interactions (GRK 1581) and is particularly grateful for having
had the opportunity to engage in this collaboration, enabled by the European
Research Council (ERC) under the Advanced Grant EFT4LHC. SR would like to thank
the Department of Science and Technology (DST), Government of India for
supporting the visit through the INSA-INSPIRE Faculty Fellowship.

\appendix

\section{Analytical and numerical results}
\subsection{Analytical results for \texorpdfstring{$Z\to Wl\nu$}{Z > Wlv}}
\label{app:calcAmpl}
The amplitude for the decay $Z\to W l \nu$ is obtained from evaluating the
diagrams shown in \autoref{fig:diagramThreeBody}. We assign the momenta
$k^\mu$, $p^\mu$ and $q^\mu$ to the final state $W$, lepton and neutrino
respectively. Allowing for modifications of the SM couplings, the amplitude 
can be written as:
\begin{multline*}
 i\mathcal A_{Z\to W l\nu} = \\-i
 \bar u(p)\left[
 \Delta^Z_{\nu}A^{\mu\nu}_1+\Delta^Z_{l}A^{\mu\nu}_2+\Delta^Z_{W}A^{\mu\nu}_3
 \right]P_L v(q)  \varepsilon^Z_\mu\varepsilon^{W\ast}_\nu\,,
\end{multline*}
with:
\begin{equation}
\begin{aligned}
A^{\mu\nu}_1 &= \frac{g^2}{\sqrt{2}c_W}(g_V^\nu+g_A^\nu) 
\gamma^\nu \frac{\slashed p +\slashed k}{(p+k)^2}\gamma^\mu\,, \\
A^{\mu\nu}_2 &= -\frac{g^2}{\sqrt{2}c_W}(g_V^l+g_A^l)
\gamma^\mu \frac{\slashed k +\slashed q}{(k+q)^2}\gamma^\nu\,, \\
A^{\mu\nu}_3 &= \frac{g^2c_W}{\sqrt{2}} \frac{V^{\mu\nu\rho}\left(p+k+q,k,p+q \right)}
{(p+q)^2-m_W^2-i\Gamma_W m_W} \gamma_\rho\,,
\end{aligned}
\end{equation}
where
\begin{equation}
\begin{aligned}
&\quad V^{\mu\nu\rho}(P,p_-,p_+)
 =g^{\mu\nu}(P+p_-)^\rho \\&\qquad
 -g^{\mu\rho}(p_++P)^\nu
 +g^{\nu\rho}(p_+-p_-)^\mu\,. 
\end{aligned}
\end{equation}
Here $m_W=80.385\,\mathrm{GeV}$ is the $W$-boson mass,
$\Gamma_W=2.085\,\mathrm{GeV}$ the $W$-boson full width and $c_W= 0.8768$ is the
cosine of the electroweak mixing angle. We extract the electroweak coupling $g$
from $g^2=8 G_F m_W^2/\sqrt{2}$, where $G_F=1.1663787\cdot
10^{-5}\,\mathrm{GeV}^{-2}$ is Fermi's constant \cite{Agashe:2014kda}. We
neglect the charged lepton mass throughout this discussion. When the functions
$A^{\mu\nu}_k$ are written in terms of the invariant masses
\begin{align}
 m_{Wl}^2&=(p+k)^2\,, & m_{l\nu}^2&=(p+q)^2\,,
\end{align}
the matrix appearing in \autoref{eq:decRate3} is defined by:
\begin{multline}
 M_{ij}=\dfrac{1}{64 (2\pi)^3 m_Z^3}
 \int \limits_{m_W^2}^{m_Z^2}		\hspace{-1ex}\dif m_{Wl}^2
 \int \limits_0^{\hat m_{l\nu}^2}	\hspace{-1ex}\dif m_{l\nu}^2
 \bigg\{\\
 	\mathrm{tr}\left[\left(A_i^{\mu\nu} \right)^\dagger \slashed p A_j^{\rho\sigma}\slashed q\right]
 	\left( \frac{1}{3}\mathcal P^Z_{\mu\rho}(p+k+q)\mathcal P^W_{\nu\sigma}(k)\right)
 \bigg\}
\end{multline}
where $\mathcal P^X_{\mu\nu}(k)=-g_{\mu\nu}+k_\mu k_\nu/m_X^2$ is the transverse 
projector for a gauge boson $X$ with momentum $k$ and mass $m_X$, $m_Z=91.19876\,
\mathrm{GeV}$ is the $Z$-boson mass and the upper integration boundary of the phase
space integral is given by:
\begin{align}
 \hat m_{l\nu}^2 = \frac{(m_{Wl}^2-m_W^2)(m_Z^2-m_{Wl}^2)}{m_{Wl}^2}\,.
\end{align}
Performing this integration numerically, and setting the width of the
$W$ to zero, leads us to the numbers shown in
\autoref{eq:decRate3}.



\subsection{Numerical results for four-body \texorpdfstring{$Z$}{Z}-decays}
\label{app:fourbody}
Here we present the numerical results of the four-body decays discuss
in the main text.

\vspace{5cm}
\begin{widetext}
\begin{eqnarray}
\frac{\Gamma(Z\to l\,l\,\nu\bar\nu)}{10^{-8}~\text{GeV}} &\simeq&
	\left\{ \begin{array}{l@{\quad}l}
		  2.8
		- 4.3 C_{\nu_l}
		+ 3.2 C_{\nu_l}^2
		- 1.3 C_{\nu_l}^3
	+ \sum_{\alpha=e,\mu,\tau} \left(
		  0.077 C_{\nu_\alpha}^2
		+ 0.27 C_{\nu_\alpha}^3
		+ 0.33 C_{\nu_\alpha}^4
	\right)
	&\text{for } l = e,\mu\\[1mm]
		  2.7
		- 4.0 C_{\nu_l}
		+ 3.0 C_{\nu_l}^2
		- 1.4 C_{\nu_l}^3
	+ \sum_{\alpha=e,\mu,\tau} \left( 
		  0.076 C_{\nu_\alpha}^2
		+ 0.26 C_{\nu_\alpha}^3
		+ 0.31 C_{\nu_\alpha}^4
	\right) 
	&\text{for } l = \tau
	\end{array}\right.
\label{llnunu}
\\[2mm]
\frac{\Gamma(Z\to jj l\,\nu_l )}{10^{-8}~\text{GeV}} &\simeq&
	\left\{ \begin{array}{l@{\quad}l}
		  8.2
		- 10 C_{\nu_l}
		+ 8.7 C_{\nu_l}^2
	&\text{for } l = e, \mu \\[1mm]
		  8.1
		- 9.9 C_{\nu_l}
		+ 8.0 C_{\nu_l}^2
	&\text{for } l = \tau
	\end{array}\right.
\label{lnujj}
\\[2mm]
\frac{\Gamma(Z\to l\:l'\nu_{l} \nu_{l'})}{10^{-8}~\text{GeV}} &\simeq&
	\left\{ \begin{array}{l@{\quad}l@{\quad}l}
	2.8 - 2.3 (C_{\nu_l} + C_{\nu_{l'}}) - 0.085 C_{\nu_l} C_{\nu_{l'}} + 1.5 (C_{\nu_l}^2 + C_{\nu_{l'}}^2 )
		&\text{for }l=e,
		&l'=\mu 
	\\[1mm]
	2.7 -  2.4 C_{\nu_l}  -  2.3 C_{\nu_{l'}}  -  0.080 C_{\nu_l} C_{\nu_{l'}} +1.5 C_{\nu_l}^2 + 1.4 C_{\nu_{l'}}^2 
		&\text{for }l = e,\mu,
		&l'=\tau
	\end{array}\right.
\label{emununu}
\end{eqnarray}
\end{widetext}

\bibliography{paper}
\bibliographystyle{apsrev4-1_title}
\end{document}